\documentclass{aa}  
\usepackage{graphicx}
\usepackage{soul}
\usepackage{txfonts}
\usepackage{upgreek}

\begin{document}

\title{The halo around HD\,32297: $\upmu$m-sized cometary dust}

\author{J. Olofsson\inst{\ref{inst:MPIA},\ref{inst:NPF},\ref{inst:IFA}}
        \and
        P. Th\'ebault\inst{\ref{inst:LESIA}}
        \and
        G. M. Kennedy\inst{\ref{inst:W1},\ref{inst:W2}}
        \and
        A. Bayo\inst{\ref{inst:ESO},\ref{inst:NPF},\ref{inst:IFA}}
      }
\institute{
    Max Planck Institut f\"ur Astronomie, K\"onigstuhl 17, 69117 Heidelberg, Germany\\\email{olofsson@mpia.de}\label{inst:MPIA}
\and
N\'ucleo Milenio Formaci\'on Planetaria - NPF, Universidad de Valpara\'iso, Av. Gran Breta\~na 1111, Valpara\'iso, Chile\label{inst:NPF}
\and
Instituto de F\'isica y Astronom\'ia, Facultad de Ciencias, Universidad de Valpara\'iso, Av. Gran Breta\~na 1111, Playa Ancha, Valpara\'iso, Chile\label{inst:IFA}
\and
LESIA-Observatoire de Paris, UPMC Univ. Paris 06, Univ. Paris-Diderot, France\label{inst:LESIA}
\and
Department of Physics, University of Warwick, Gibbet Hill Road, Coventry, CV4 7AL, UK\label{inst:W1}
\and
Centre for Exoplanets and Habitability, University of Warwick, Gibbet Hill Road, Coventry CV4 7AL, UK\label{inst:W2}
\and
European Southern Observatory, Karl-Schwarzschild-Strasse 2, 85748 Garching bei M\"unchen, Germany\label{inst:ESO}
}


\abstract{The optical properties of the second generation dust that we observe in debris disks remain quite elusive, whether it is the absorption efficiencies at millimeter wavelengths or the (un)polarized phase function at near-infrared wavelengths. Thankfully the same particles are experiencing forces that are size dependent (e.g., radiation pressure), and with high angular resolution observations we can take advantage of this natural spatial segregation.}
{Observations at different wavelengths probe different ranges of sizes; millimeter observations trace the larger grains while near-infrared observations are sensitive to the other extreme of the size distribution. There is therefore a great synergy in combining both observational techniques to better constrain the optical properties of the particles.}
{We present a new approach to simultaneously model SPHERE and ALMA observations and apply it to the debris disk around HD\,32297, putting the emphasis on the spatial distribution of the grains with different $\beta$ values. This modeling approach requires few assumptions on the actual sizes of the particles and the interpretation can therefore be done a posteriori.}
{We find that the ALMA observations are best reproduced with a combination of small and large $\beta$ values ($0.03$ and $0.42$) while the SPHERE observations require several intervals of $\beta$ values. We discuss the nature of the halo previously reported in ALMA observations, and hypothesize it could be caused by over-abundant $\mu$m-sized particles (the over-abundance being the consequence of their extended lifetime). We model the polarized phase function at near-infrared wavelengths and fluffy aggregates larger than a few $\mu$m provide the best solution.}
{Comparing our results with comets of the solar system, we postulate that the particles released in the disk originate from rather pristine cometary bodies (to avoid compaction of the fluffy aggregates) and are then set on highly eccentric orbits, which could explain the halo detected at long wavelengths.}

\keywords{Stars: individual (HD\,32297) -- circumstellar matter -- Techniques: high angular resolution -- polarimetry}
\maketitle
%

\section{Introduction}

The second generation dust that we observe in debris disks is continuously replenished from a collisional cascade of large planetesimals (see \citealp{Krivov2010}, \citealp{Hughes2018}), or, in some rarer cases, by transient events such as the violent breakup of larger planetary embryos (\citealp{Jackson2014}, \citealp{KRal2015}). The physical properties of the particles released from such collisions still remain quite elusive, even though it is becoming clear that the assumption of compact spherical grains fails to reproduce contemporary observations (e.g., \citealp{Milli2017,Milli2019}). Debris disks are interesting targets to try and better characterize the dust optical properties for two reasons. First, they are optically thin at all wavelengths and we do not have to account for multiple scattering events or estimating the temperature in an optically thick medium, simplifying (to some extent) the modeling of the observations. Second, the dynamics of the particles strongly depend on their sizes. For most stellar spectral types, radiation pressure, which can be parametrized by the unitless ratio $\beta$ between radiation pressure and gravitational forces ($\beta \propto 1/s$ for grains larger than a few $\mu$m), will naturally result in different spatial extent for different grain sizes\footnote{If the disk is not entirely free of gas, the drag it exerts on the grains is also a size-dependent force, see for instance \citet{Takeuchi2001}}. The smaller particles will be set on highly eccentric orbits while larger ones will remain on orbits very similar to the ones of the parent bodies. Since observations at different wavelengths probe different grain sizes, we can take advantage of this natural spatial segregation (\citealp{Thebault2014}).

Millimeter (mm) observations inform us about the spatial distribution of the large dust particles, and therefore best trace the location of the birth ring of planetesimals where the collisions are taking place. On the other hand, near-infrared (IR) scattered and polarized light observations are sensitive to the other extreme of the size distribution and trace $\mu$m-sized dust grains. Consequently, there is a great synergy between the two observational techniques; by constraining the location of the birth ring from mm observations, we can then have a better description of where the small dust grains should be launched from on highly eccentric orbits. This leads to a more accurate description of the spatial distribution of ``observable'' dust particles (with sizes $s \lesssim 1$\,mm), which can be used to better constrain their optical properties, such as the absorption efficiencies and near-IR phase function.

The debris disk around the A0 star HD\,32297 is an ideal candidate to investigate the spatial distribution of second generation dust using multi-wavelengths observations. It has been spatially resolved in the near-IR (e.g., \citealp{Kalas2005}, \citealp{Rodigas2014}, \citealp{Schneider2014}, \citealp{Bhowmik2019}, \citealp{Esposito2020}). The disk is seen almost perfectly edge-on, and displays extended swept-back wings (best seen in the \textit{Hubble Space Telescope} observations of \citealp{Schneider2014}). The disk has also been observed at mm wavelengths, and spatially resolved along the major-axis (\citealp{MacGregor2018}, \citealp{Cataldi2020}). Interestingly, \citet{MacGregor2018} reported the presence of a halo in the ALMA Band\,6 ($1.3$\,mm) dataset and concluded that this additional component cannot reasonably arise from $\mu$m-sized dust grains. Nonetheless, the detection of such a halo, possibly composed of mm-sized particles, is surprising as these grains should in principle not venture very far away from the birth ring, justifying further investigation using multi-wavelength observations. The almost edge-on configuration of the disk does come at a cost since information is lost due to projection effects, however, highly inclined disks allow for a wider range of scattering angles to be probed to constrain the phase function. Furthermore, due to the same projection effects, extended halos are easier to detect at all wavelengths as we probe larger column densities at all distances. HD\,32297, having high angular resolution observations at both near-IR and sub-mm wavelengths and displaying a rather unique halo in ALMA observations, is therefore an interesting target to further study.

In the literature, observations (either near-IR or mm) are usually reproduced using geometric models that do not necessarily include any physics in them, and the interpretation of the physical processes at play is done a posteriori (\citealp{Augereau1999}, \citealp{Marino2016}, \citealp{Engler2017}, \citealp{Kennedy2018}, \citealp{Daley2019}, \citealp{Matra2019}, \citealp{Milli2019}, \citealp{Ren2019}, \citealp{Olofsson2020}, among others). In some cases, more complex models are used to model the observations, taking for instance into account the effect of radiation pressure or stellar winds on the small particles (e.g., \citealp{Esposito2016}, \citealp{Olofsson2019}, \citealp{Adam2021}), or the intrinsic width of the parent planetesimal belt (\citealp{Kennedy2020}). We here propose a novel approach at modeling near-IR and sub-mm observations, belonging to this second ``family'' of models. This approach can provide more stringent constraints on the properties of the dust particles. The philosophy is to identify the spatial scales, which are intrinsically related to typical grain sizes, that are most representative of the observations. By computing images for different $\beta$ intervals, we can then identify the intervals (i.e., spatial extent) best suited to match the observations. This information can then be used to derive the properties of the dust grains (e.g., absorption efficiencies, phase function, porosity). In this paper, we first describe the multi-wavelength observations of the disk around HD\,32297, how we model them, and discuss the results, before concluding.

\section{Observations}\label{sec:obs}

\subsection{SPHERE observations}

To probe the population of small dust particles, we used the ``Spectro-Polarimetric High Contrast Exoplanet REsearch'' (SPHERE, \citealp{Beuzit2019}) dual-beam polarimetric imaging (DPI, \citealp{Dohlen2008}) $J$-band observations of HD\,32297, which were first presented in \citet{Bhowmik2019}. We reduced the data using the \texttt{IRDAP}\footnote{Available at \url{https://irdap.readthedocs.io/en/latest/}} package presented in \citet[][version 1.3.1]{vanHolstein2020}. The outputs of the pipeline are the $Q_\phi$ and $U_\phi$ images, the former containing the polarized signal from the disk (left panel of Fig.\,\ref{fig:sphere}), while the latter is free of astrophysical signal (assuming single scattering events) and is used as a proxy for the uncertainties.

\subsection{ALMA observations}

To constrain the location of the birth ring, we used the Band 8 ($615$\,$\mu$m) ALMA observations published in \citet[][program ID 2017.1.00201.S]{Cataldi2020}, with naturally weighted beam size of $0.63\arcsec \times 0.54\arcsec$ (marginally smaller than the observations presented in \citealp{MacGregor2018} with a beam of $0.76\arcsec \times 0.51\arcsec$). The data was reduced using the ``Common Astronomy Software Applications'' package (\texttt{CASA}, version 5.6, \citealp{CASA}). Once the reduction was performed using the script provided by the observatory, we re-evaluated the weights ($\propto 1/\sigma^2$) using the \texttt{statwt} task, and used the \texttt{split} task to average the complex visibilities to a single channel per spectral window, and a time bin of $30$\,sec, before exporting them using the \texttt{uvplot} package (\citealp{uvplot_mtazzari}). The bottom left panel of Figure\,\ref{fig:sphere} shows the observations (using \texttt{tclean} and natural weighting) and the beam size in the lower left corner. The noise level was estimated to $\sigma = 0.3$\,mJy/beam, from the clean image, in regions where there is no disk signal.

\section{Geometrical fit to the observations}\label{sec:code}

As mentioned before, the outward extent of the disk is governed by the radiation pressure, parametrized by the $\beta$ ratio. By ``launching'' thousands of particles with different $\beta$ values, similarly to \citet{Lee2016}, we can therefore probe different spatial scales. Furthermore, by working with $\beta$ values, we can make minimal assumptions on the exact sizes of the dust particles or their composition. Our goal is to identify which spatial regions contribute the most to best match the observations, taking into account that particles with different $\beta$ values probe different spatial scales. 

\subsection{The dynamics of dust particles}

For a given model\footnote{The code is available at \url{https://github.com/joolof/betadisk}}, we ``launch'' $n_\mathrm{dust}$ particles with different $\beta$ values (see next paragraph), representing dust grains of various sizes, assuming that their parent bodies have a semi-major axis $a$, eccentricity $e$, and argument of periapsis $\omega$. The semi-major axis $a$ is drawn from a normal distribution\footnote{We therefore do not consider the possible presence of gaps or that the disk is highly structured.} centered at $a_0$ with a standard deviation $\delta_\mathrm{a}$. The launch point is uniformly distributed in mean anomaly between $-\pi$ and $\pi$, and is then converted into a true anomaly $\nu$ by solving the Kepler equation for $e$. The initial velocity of the particle is assumed to be the Keplerian velocity of the parent body. Following, \citet{Wyatt1999,Wyatt2006}, and \citet{Lee2016}, for each particle (a given $\beta$ and $\nu$), the new orbital elements are computed as
\begin{equation}\label{eqn:orbit}
\begin{aligned}
a_{\mathrm{n}} = \cfrac{a (1 - \beta)}{1 - 2\beta \cfrac{1 + e \mathrm{cos}(\nu)}{1 - e^2} }, \\
e_{\mathrm{n}} = \frac{1}{1 - \beta} \times \sqrt{e^2 + 2\beta e \mathrm{cos}(\nu) + \beta^2}, \\
\omega_{\mathrm{n}} = \omega + \mathrm{arctan}\left[\frac{\beta \mathrm{sin}(\nu)}{\beta\mathrm{cos}(\nu) + e}\right],
\end{aligned}
\end{equation}
For each particle, we draw a longitude of ascending nodes uniformly between $-\pi$ and $\pi$ and an inclination following a normal distribution centered at $0$\,radians, with a standard deviation $\psi$. We then draw a value for the mean anomaly ($\in ]-\pi, \pi]$) and compute the corresponding true anomaly (using the eccentricity $e_\mathrm{n}$). The $(x, y, z)$ positions of the particle, as well as its distance to the star $r = \sqrt{x^2 + y^2 + z^2}$, are then calculated using all the orbital parameters. To limit the number of free parameters, we set the eccentricity of the parent bodies to $e = 0$ (in that case, the value of $\omega$ no longer matters, removing another free parameter). The disk around HD\,32297 is not known for showing any signs of eccentricity and the edge-on configuration is the most challenging to properly constrain this parameter. Implicitly, we are making the assumption that all the parent bodies have small eccentricities (to ensure enough collisions are taking place), but that on average, the parent belt is circular.

We consider $n_\mathrm{dust}$ particles, with $\beta$ values between $\beta_\mathrm{min}$ and $\beta_\mathrm{max}$. To account for the fact that particles on highly eccentric orbits will spend most of their time in low density regions, and therefore survive longer, we also compute the following ``correction factor'' $\alpha = (1 - \beta)^{3/2} / [1 - e^2 - 2 \beta \times (1 + e\mathrm{cos}(\nu))]^{3/2}$ for each particle (\citealp{Strubbe2006}, \citealp{Thebault2008}). Up until now, the free parameters are $a_0$, $\delta_\mathrm{a}$, and $\psi$.

\subsection{Synthetic images}

To create images to be compared with the observations, there are three additional parameters, the pixel size, the inclination of the disk $i$ and its position angle $\phi$. Each of the $(x, y, z)$ position are projected and rotated to account for $i$ and $\phi$, respectively. We then find the image pixel closest to the new values, accounting for the pixel size.

For scattered light images, we compute the scattering angle $\theta$ (the angle between the star, the particle, and the observer) using the dot product between a unit vector along the line of sight and the rotated 3D coordinates of the particle. The contribution of a particle to a pixel of the image will then be $\alpha S_{12}(\theta) / (r \beta)^2$, where $S_{12}$ is the polarized phase function (see later), and $\theta$ the scattering angle. The $1/r^2$ accounts for the illumination factor, and the $1/\beta^2$ represents the cross section of the particle.

For mm images, from the distance $r$ (in au), we estimate the size-independent temperature of the grains $T_\mathrm{dust}$ as $T_\mathrm{dust} = 278.3\,L_\star^{1/4} / \sqrt{r}$ (\citealp{Wyatt2008}), compute the Planck function $B_\nu(\lambda, T_\mathrm{dust})$ at the wavelength of interest $\lambda$, and the contribution of a single particle will be $\alpha B_\nu(\lambda, T_\mathrm{dust})/\beta^2$, therefore assuming that the particles behave as perfect blackbodies (the $1/\beta^2$ accounts for the surface area of the particles) .

\subsection{Words of caution}

The main benefit of the approach described above is that it relies as little as possible on the optical properties of the dust particles, therefore decreasing the number of free parameters, such as the dust composition, the porosity, or the light scattering theory used. While this speeds up the computational time, this also comes at a cost. For instance, both the phase function $S_\mathrm{12}$ and the temperature $T_\mathrm{dust}$ should in fact depend on the size of the particles, and the contribution to the thermal emission (polarized scattered light) should also depend on the absorption (scattering) efficiencies of the particles. 

\subsection{Modeling strategy}

To find the best fit solution we used the \texttt{emcee} affine-invariant ensemble sampler (\citealp{emcee}), with the following four free parameters: $a_0$, $ \delta_a$, and two standard deviations $\psi_\mathrm{SPHERE}$, and $\psi_\mathrm{ALMA}$, for the near-IR and mm observations, respectively. The distribution of $\beta$ values will be evaluated independently for each realization of the \texttt{emcee} sampler (see later), which is parametrized to use $20$ ``walkers'' and a chain length of $10,000$ for each of them. The need for two different scale heights for the SPHERE and ALMA data is motivated by the work presented in \citet{Olofsson2022} where they showed that the scale height can vary as a function of the wavelength (disks being flatter at near-IR wavelengths compared to mm wavelengths) if the disk harbors gas, which is the case for HD\,32297 (\citealp{Greaves2016}, \citealp{MacGregor2018}, \citealp{Cataldi2020}). Vertical stratification is also expected when the disk is devoid of gas (\citealp{Thebault2009}), but in this case the disk should be seen flatter at mm wavelengths compared to near-IR observations. The inclination and position angle of the disk are not free parameters in the modeling, since the ALMA observations do not have a comparable angular resolution compared to the SPHERE ones. Furthermore, \citet{Cataldi2020} noted a possible discrepancy when comparing the inclination derived from near-IR scattered light observations and the \ion{C}{I} emission line. The authors found a smaller inclination ($i \sim 77.9^{\circ}$) in the latter case, but the origin of this discrepancy is not clear. We therefore opted to fix both parameters to $i = 86.9^{\circ}$ and $\phi = -132.3^{\circ}$ (from the modeling of the same SPHERE observations, presented in \citealp{Olofsson2022}), thus reducing the amount of free parameters to four.

The other (fixed) parameters required for the fitting are the following; $L_\star = 7.61$\,$L_\odot$ (\citealp{Olofsson2022}), $d_\star = 129.71$\,pc (\citealp{Gaia2016,Gaia2020}), $n_\mathrm{dust} = 10\,000\,000$, $n_\beta = 10$,  and $\lambda = 1.22$ and $615$\,$\mu$m (SPHERE and ALMA, respectively). $\beta_\mathrm{min}$ and $\beta_\mathrm{max}$ are set to $0.01$ and $0.49$, respectively, to capture a wide range of spatial scales. As mentioned in \citet{Strubbe2006}, the contribution of unbound particles to the surface brightness profiles should be marginal at best, and we therefore opted for a cut-off at $0.49$.  The pixel scales for the synthetic images are $12.26$ and $5$\,mas for the SPHERE and ALMA observations, respectively.

During the fitting process, the sampler will draw values for each of the four free parameters, and synthetic images will be computed for both SPHERE and ALMA. Since our goal is to identify the spatial scales that are most representative of the observations, rather than creating one single image by summing over the whole range of $\beta$, we instead create $n_\beta$ images, sampling $n_\beta$ intervals of $\beta$ values (linearly spaced between $\beta_\mathrm{min}$ and $\beta_\mathrm{max}$). For each set of free parameters ($a_0$, $\delta_a$, and both $\psi$ values), we then find the linear combination of these $n_\mathrm{\beta}$ images that best reproduces the observations (using the \texttt{lmfit} package, \citealp{lmfit}, which also provides the confidence interval for each weight). In practice, this means that we are relaxing the assumption of a fixed grain size distribution from the ``cross-section'' model, and are re-evaluating the contributions of each $\beta$ interval. 

For the ALMA observations, the \texttt{lmfit} modeling is performed in the $u-v$ space. Since the Fourier transform is a linear operation, we can compute the visibilities for each $n_\beta$ images and find the best combination to fit the observations by scaling them. Following \citet{Cataldi2020}, when computing the images, we also included an offset to the model, of $\delta_\mathrm{RA} = -0.06\arcsec$ and $\delta_\mathrm{Dec} = 0.04\arcsec$.

For the SPHERE observations, we still need to account for the polarized phase function. Instead of using a parametrized one (e.g., Henyein-Greenstein), we follow the approach outlined in \citet{Olofsson2020}, where the phase function is an output of the modeling process (and therefore does not depend on the optical properties of the particles). For a given set of free parameters, we run a first model with an isotropic without any polarized phase function and find the best linear combination of the $n_\beta$ frames using \texttt{lmfit}. Since the phase function is not included, the resulting image will therefore trace the dust density distribution. For both the observations and the model, we then compute the scattering angle for each pixel, and estimate the brightness profile as a function of $\theta$. The phase function that will best reproduce the observations is estimated as the brightness profile of the observations divided by the one of the model (as detailed in \citealp{Olofsson2020} only one iteration is required to derive the most suitable phase function). We then compute a new model, this time using the phase function that was just estimated for all the $n_\beta$ images. We again find the linear combination of those images minimizing the differences compared to the SPHERE DPI observations.

Once a model has been computed for both the SPHERE and ALMA observations, we compute the joint goodness of fit. Because the ALMA and SPHERE observations do not have the same degrees of freedom\footnote{For the SPHERE observations, the $\chi^2$ is evaluated within the dashed ellipse shown on the upper left panel of Fig.\,\ref{fig:sphere}, and the central mask is ignored.}, one dataset may dominate the total $\chi^2$ over the other. Therefore, prior to starting the modeling, for both datasets we first compute the $\chi^2$ for the null hypothesis, and compute a scaling factor so that the largest of the two values is equal to the other one. In practice, we multiply the $\chi^2$ from the ALMA observations by a factor $0.22$. Finally, the likelyhood for this realization of free parameters is returned to the \texttt{emcee} sampler, and another set of free parameters can be drawn. It would have been ideal to avoid the two-fold approach (\texttt{emcee} combined with \texttt{lmfit}) and include the re-weighting of the size distribution as free parameters in the \texttt{emcee} fitting. Unfortunately, computing one model is quite expensive, and ensuring the convergence of the fitting process over a $14$ dimension space would be too costly.

\section{Results and discussion}\label{sec:discussion}

\subsection{Disk geometry}

\begin{figure*}
\centering
\includegraphics[width=\hsize]{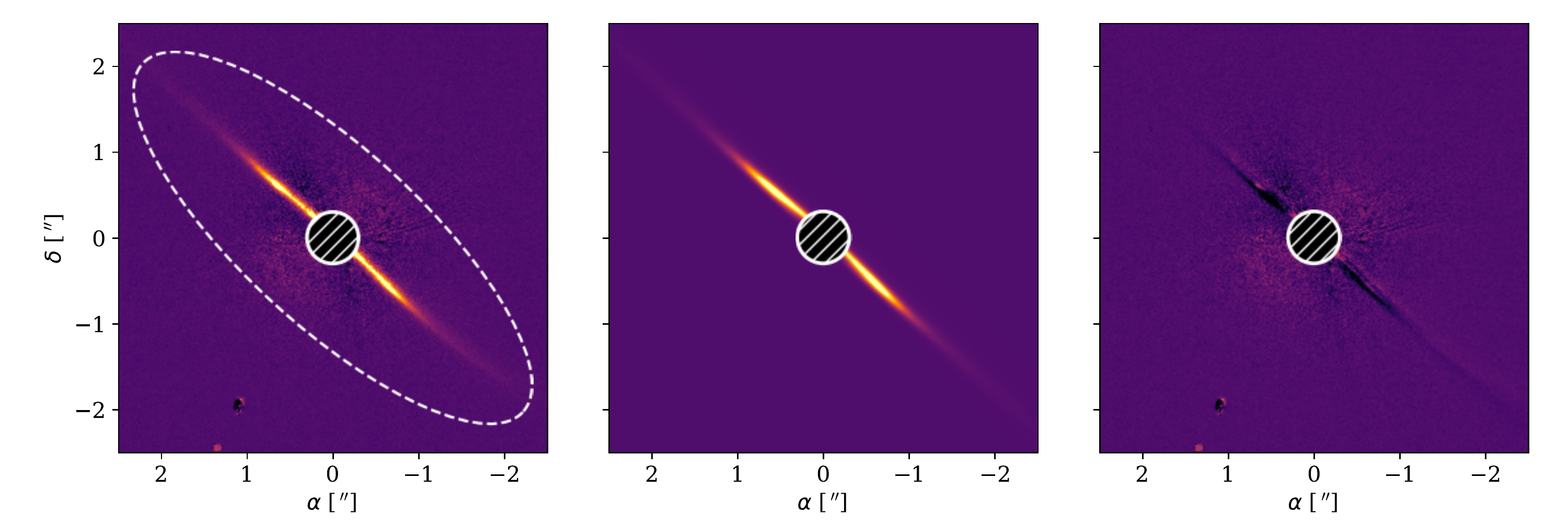}
\includegraphics[width=\hsize]{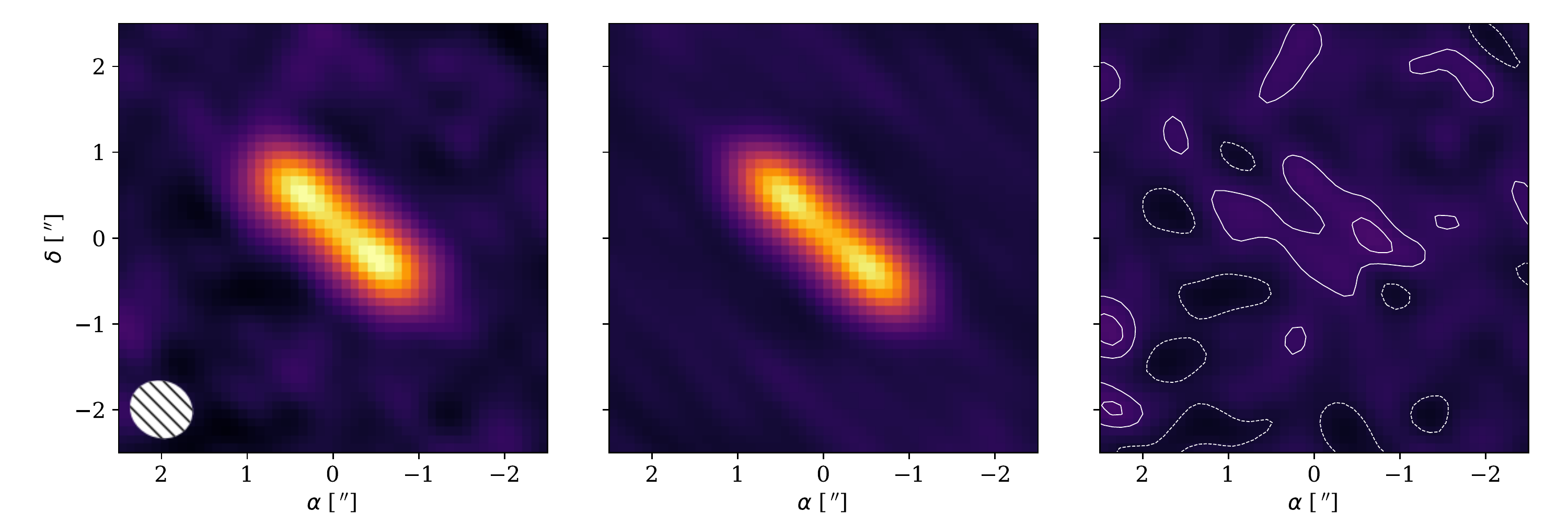}
    \caption{From left to right: observations, best fit model, and residuals, for the SPHERE and ALMA observations (top and bottom, respectively). The scaling is linear and the same for all horizontal panels. For the ALMA observations, the beam is shown in the leftmost panel, and the contours on the right panel are at $[-1, 1, 2]\,\sigma$ ($\sigma = 0.3$\,mJy/beam).}
\label{fig:sphere}
\end{figure*}

\begin{table}
\caption{Details for modeling of observations and best-fit results.}
\label{tab:grid}
\centering
\begin{tabular}{lcc}
\hline\hline
Parameters & Prior & Best-fit \\
\hline
$a_0$ [$^{\arcsec}$] & [0.5,1.5] & $0.99_{-0.05}^{+0.03}$\\
$\delta_a$ [$^{\arcsec}$] & [0.01,0.5] & $0.22_{-0.01}^{+0.01}$\\
$\psi_\mathrm{SPHERE}$ [rad] & [0.001,0.1] & $0.005 (\leq 0.014$)\\
$\psi_\mathrm{ALMA}$ [rad] & [0.005,0.1] & $0.007 (\leq 0.071$)\\
\hline
\end{tabular}
\end{table}

Table\,\ref{tab:grid} lists the free parameters, the uniform linear priors, as well as the best-fit results along with their uncertainties. The density distributions are displayed in Fig.\,\ref{fig:corner_obs}, using the \texttt{corner} package (\citealp{corner}) and show that while both $a_0$ and $\delta_a$ are well constrained, the probability density distributions for the two scale heights $\psi_\mathrm{SPHERE}$ and $\psi_\mathrm{ALMA}$ are most likely upper limits, meaning that the disk is not vertically resolved in either dataset (as already mentioned in \citealp{Olofsson2022} regarding the SPHERE observations). For these two parameters, for the best fit model we took the value at the peak of the distributions ($\psi_\mathrm{SPHERE} = 0.005$ and $\psi_\mathrm{ALMA} = 0.007$\,rad). For the confidence interval, we estimated the bin of the histogram for which the integral of the distribution up to that bin reaches $84$\% of the total integral ($\leq 0.014$ and $\leq 0.071$, respectively). We find that the birth ring of the disk is best described by a normal profile centered at $\sim 1$\arcsec ($129.7$\,au) with a standard deviation of $0.22$\arcsec ($28.5$\,au), in agreement with the modeling results of \citet{MacGregor2018}, \citet{Bhowmik2019}, and \citet{Cataldi2020}. From $a_0$ and the full width at half maximum of the radial profile, we find a fractional width ($\mathrm{FWHM}/a_0$) of $0.52$. Even though the almost edge-on inclination is not ideal to constrain the width of the disk, this is in good agreement with other disks observed with ALMA (median width of $0.74$, \citealp{Marino2021,Marino2022}). When summing the total flux in the best-fit model for the ALMA observations, we derive a total flux of $23.4\pm0.3$\,mJy, compatible with the value reported in \citet[][$22.0 \pm 2$\,mJy]{Cataldi2020}.

The upper panels of Figure\,\ref{fig:sphere} show the observations, best-fit model, and residuals for the SPHERE DPI observations. Most of the signal from the disk is removed, though some residuals remain, both positive and negative along the major axis of the disk (similarly to \citealp{Duchene2020} and \citealp{Olofsson2022}). The lower panels of the same Figure show the results for the ALMA observations. To convolve the best-fit model, we replaced the observed visibilities by the ones of the model, and used \texttt{tclean} to compute the image. For the residuals, we proceeded the same way, but this time subtracting the complex visibilities of the model to the observed ones\footnote{Using \texttt{Python} methods provided here \url{https://github.com/drgmk/alma}}. The contours on the residuals are $[-1, 1, 2]\,\sigma$ (and there is nothing above $3\,\sigma$). There is still some signal, at the $1-1.5\,\sigma$ level above and below the major axis of the disk, which most likely explains why \citet{Cataldi2020} derived a smaller inclination when modeling the observations.

\subsection{The nature of the mm halo}\label{sec:halo}

This sub-section focuses on the halo that was reported in ALMA Band\,6 observations by \citet{MacGregor2018}. We will first demonstrate that the halo is also detected in the Band\,8 observations, before discussing whether is it composed of mm- or $\mu$m-sized dust grains. Afterwards, we discuss what this implies for the optical dust properties and their emissivity at long wavelengths, and discuss its possible origin and detectability.

\subsubsection{Recovering the halo in Band 8}

\begin{figure}
\centering
\includegraphics[width=\hsize]{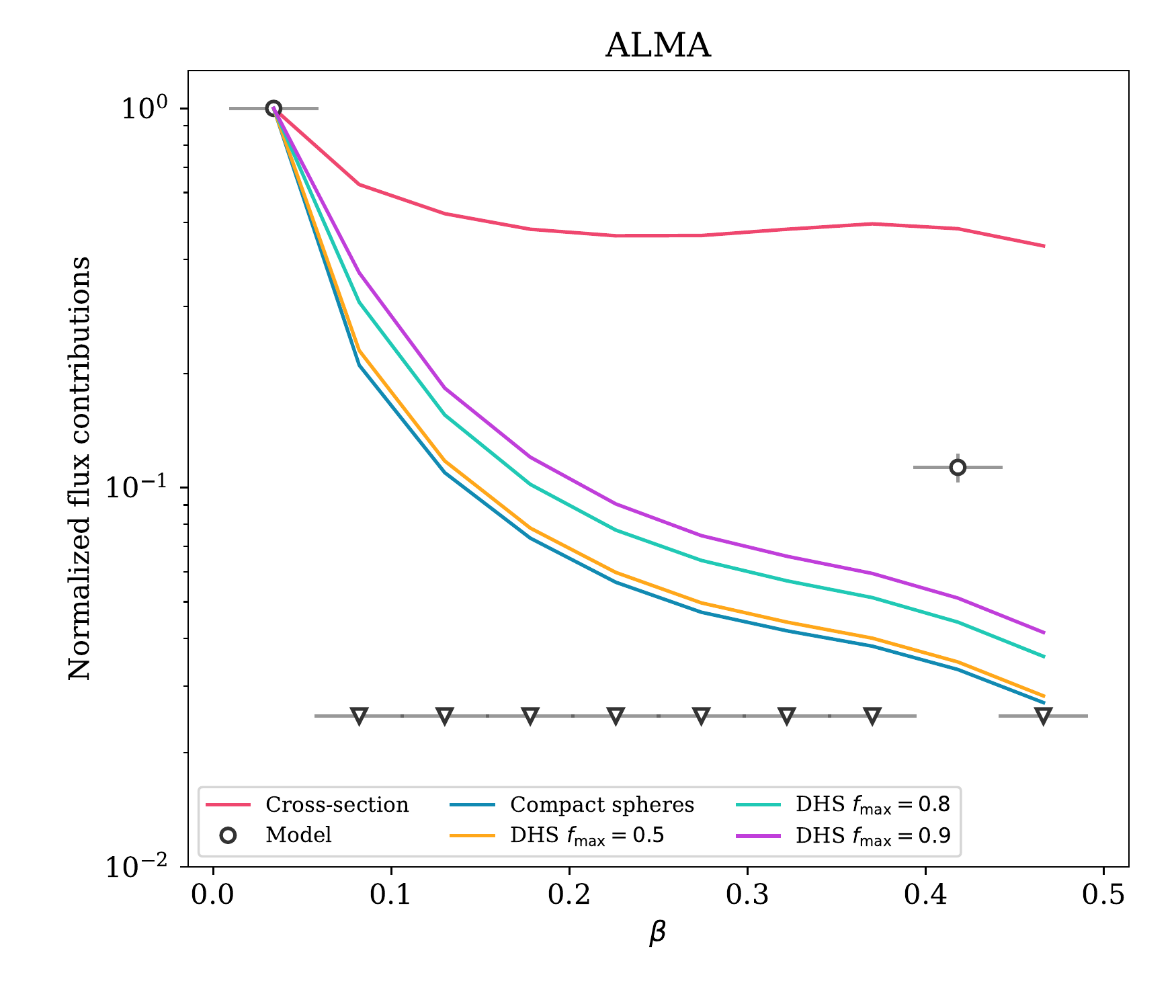}
    \caption{Normalized flux contributions as a function of $\beta$ for the ALMA observations. The red line shows the contributions derived from the ``cross-section'' model. The open symbols show the contributions of the $\beta$ intervals for the best fit model when relaxing the assumption on the grain size distribution. The values that are below $2.5 \times 10^{-2}$ are considered as upper limits, and are represented by downward triangles. The horizontal bars show the width of the $\beta$ intervals, and the vertical bars are the $1\sigma$ confidence interval. The other solid lines show the contributions for models that account for the optical properties of the particles (see text for details).}
\label{fig:contrib_alma}
\end{figure}

Before further describing the results of our modeling, we compute a reference model using the best-fit parameters listed in Table\,\ref{tab:grid}. But for this model, instead of evaluating the weights of each $\beta$ intervals, the size distribution is fixed to the canonical differential power-law $\mathrm{d}n(s) \propto s^{-3.5}\mathrm{d}s$ (assuming $\beta \propto s^{-1}$ it is equivalent to $\mathrm{d} n(\beta) \propto \beta^{3/2} \mathrm{d}\beta$, \citealp{Lee2016}). This model, referred to as the ``cross-section'' model since it should account for most of the effects (size, distribution, cross section, extended lifetime), will serve as a guide to interpret our findings. Figure\,\ref{fig:contrib_alma} shows the relative contributions to the thermal emission in the sub-mm for each of the $n_\mathrm{\beta}$ images, normalized to their maximum. We first focus on the red line which shows the flux contribution of each $\beta$ interval obtained for this ``cross-section'' model, which does not take into account the optical properties of the dust grains and has a fixed grain size distribution (the other solid lines will be discussed later on). It should first be noted that this profile remains relatively flat for $\beta \gtrsim 0.2$ despite the apocenter of those grains being farther and farther away from the star and their temperature thus becoming lower. For the integrated flux, this effect is counter-balanced by \textit{(i)} the top-heavy size distribution and \textit{(ii)} the correction factor $\alpha$, accounting for these grains' extended lifetime. The combination of those two factors results in a flux multiplied by a factor $[\beta(1-\beta)/(1-2\beta)]^{3/2}$, which increases significantly for large $\beta$ values.

The results when relaxing the constraints on the size distribution (linear combination obtained using \texttt{lmfit}, instead of imposing the size distribution) are shown as open black symbols in Figure\,\ref{fig:contrib_alma}. These symbols show the flux contributions required to reproduce the ALMA observations as a function of $\beta$ (contributions below $2.5 \times 10^{-2}$ are represented by downward triangles and are considered upper limits). The ALMA observations can be best reproduced using only two intervals of $\beta$ values, $\beta_0 = 0.034$ and $\beta_8 = 0.42$, these two intervals contributing to $\sim 90\pm4$ and $\sim 10\pm1$\% of the total flux, respectively. While it is not surprising that the total flux is dominated by the main belt of the disk (large grains having low $\beta$ values), the observations are best explained when also considering particles with large $\beta$ values. Since those particles have large eccentricities, they contribute to a spatially extended emission, whose contribution is not negligible ($\sim 10$\%). This requirement is purely driven by the radial extent of the disk. This means that the halo reported in \citet{MacGregor2018} is therefore also detected in the Band\,8 observations. 

\subsubsection{Mm- or $\upmu$m-sized particles?}

Up until now, we only made very few assumptions that directly relate $\beta$ to the grain size $s$. The only occasions where we needed to use the grain size were when computing the images as we need the cross-section and surface area of the grains (for scattered light and mm observations, respectively). If we want to further discuss the properties of the dust grains, we need to relate the $\beta$ values with typical sizes $s$. For instance, up until now, we cannot discard the possibility that the high-beta particles contributing to $\sim$10\% of the total flux are in fact large mm-sized grains that acquired high eccentricities by another mechanism than radiation pressure. 
If mm-sized grains are responsible for the halo (particles with $\beta = 0.42$), our modeling results suggest that they need to have an eccentricity of $\beta_8 / (1 - \beta_8) \sim 0.72$ and a semi-major axis $(1-\beta_8)/(1 - 2 \beta_8) \sim 3.5$ times larger than the semi-major axis of the parent body they were released from. Since we find the parent belt to be at $1\arcsec$, corresponding to $129.7$\,au, this translates into an apocenter distance of $\sim 800$\,au for the mm-sized particles. Overall, this begs the question of whether there is a physical mechanism able to excite the eccentricities of large mm-grains to such high values?

As mentioned previously, the presence of CO and \ion{C}{I} gas has been reported for the disk around HD\,32297, and \citet{MacGregor2018} suggested that gas drag could be a possible mechanism to alter the dynamics of the large grains. Indeed, \citet{Olofsson2022} showed that gas drag can have an impact on the radial distribution of particles in debris disks. However, they also showed that large grains are largely unaffected by the gas drag force and do not migrate outward. As a matter of fact, large particles are expected to migrate inward (see also \citealp{Krivov2009}). Alternative mechanisms mentioned in \citet{MacGregor2018} could be interactions with the interstellar medium or planet-disk interactions. However, as discussed by the authors, the former should predominantly act on small particles, while the latter has mostly been studied for scattered light observations (e.g., \citealp{Thebault2012}, \citealp{Lee2016}) and not for large grains. It is therefore quite challenging to identify a mechanism that could excite the eccentricity of mm-sized grains up to $\sim 0.7$. Consequently, following the principle of Occam's razor, we will now assume that large $\beta$ values do correspond to small dust particles, as radiation pressure can naturally and efficiently increase the eccentricity of those grains.

\subsubsection{Optical properties of the dust particles}

The assumption that the halo that is observed at mm wavelength is composed of small particles is however not free of challenges, as noted by \citet{MacGregor2018}, as small dust grains are usually poor emitters at long wavelengths. To further investigate this, we used the \texttt{optool}\footnote{Available at \url{https://github.com/cdominik/optool}} package (\citealp{Dominik2021}, \citealp{Toon1981}) to compute the absorption and scattering efficiencies ($Q_\mathrm{abs}$ and $Q_\mathrm{sca}$, respectively), and the asymmetry parameter $g_\mathrm{sca}$ for grains of different sizes. We used a mixture of amorphous silicates (optical constants from \citealp{Dorschner1995}) and amorphous carbon (\citealp{Zubko1996}). The mixing is done with volume fractions of $60$\% amorphous silicates, $15$\% amorphous carbon, and $25$\% porosity (the standard ``DIANA'' setup, \citealp{Woitke2016}, commonly used for circumstellar disks). When computing the optical properties of dust particles, we used the Mie theory (\citealp{Mie1908}) for compact spheres, and the ``Distribution of Hollow Spheres'' (DHS, \citealp{Min2005}) to mimic irregularly shaped grains, with different values for the maximum filling factor $f_\mathrm{max}$ (to parametrize the departure from spherical grains). We used \texttt{optool} to compute the properties for a single grain size $s$, and compute the corresponding $\beta$ value as
\begin{equation}\label{eqn:beta}
    \beta(s) = \frac{3 L_\star}{16 \pi G c^2 M_\star} \times \frac{Q_{\mathrm{pr}}(s)}{\rho s},
\end{equation}
where $L_\star$ and $M_\star$ are the stellar luminosity and mass ($7.61$\,$L_\odot$ and $1.78$\,$M_\odot$, respectively), $G$ the gravitational constant, and $\rho$ the dust density (evaluated by \texttt{optool} depending on the input parameters). The radiation pressure efficiency $Q_\mathrm{pr}$ is equal to $Q_\mathrm{abs} + (1 - g_\mathrm{sca}) Q_\mathrm{sca}$ averaged over the stellar spectrum. Since the relationship between $\beta$ and $s$ will depend on the shape of the grains, we sampled a range of sizes large enough to ensure that it will cover the range of $\beta$ between $0$ and $0.5$ after converting $s$ to $\beta$ using Eqn.\,\ref{eqn:beta}. Afterwards, we interpolated the $Q_\mathrm{abs}$ values for the values of $\beta$ used in the modeling. The colored lines (other than the red one) in Figure\,\ref{fig:contrib_alma} show the expected flux distribution from the ``cross-section'' model (which assumes a size distribution in $s^{-3.5}$) multiplied by the absorption efficiencies, the only ingredient that was missing from the previous analysis. The first thing to be noted is that irregularly shaped grains are overall more efficient than compact spheres at contributing to the thermal emission as discussed in \citet{Min2016} and \citet{Tazaki2018}. Even though small grains are not efficient emitters at mm wavelengths, considering grains with a maximum filling factor $f_\mathrm{max} = 0.8$ helps increase their contribution to the mm flux, compared to spherical grains.

The second interesting result is that the expected flux distribution of the ``cross-section'' model (red solid line) clearly over-predicts the flux required to reproduce the halo (interval corresponding to $\beta = 0.42$) by a factor $\sim 4$, but that when accounting for the values of $Q_\mathrm{abs}$, those models under-predict the required flux by a factor $\sim 2-3$ (other colored solid lines). Ideally, this could be used to better constrain the values of $Q_\mathrm{abs}$ of particles of different size\footnote{If indeed the assumed size distribution in $s^{-3.5}$ is correct. An alternative explanation could be an over-abundance of small dust particles.}. Indeed, the cross-section model accounts for most of the effects, the size distribution, the correction factor, the surface area, and (at first order) the temperature of the grains. Therefore, dividing the flux contributions needed to model the observations by the contributions expected from the cross-section model should in principle yield a robust estimate of the $Q_\mathrm{abs}$ values as a function of $\beta$. Unfortunately, this is not straightforward, since we do not have a continuous distribution of fluxes as the output of our modeling strategy. The fitter ``prefers'' to use two discrete bins of $\beta$ rather than a continuous distribution as a function of $\beta$. It may be the case that the contribution of particles with $\beta$ values in the range $[0.3, 0.5]$ is all accounted for in the single $\beta_8$ interval, thus artificially increasing its contribution to the thermal emission. A possible explanation being that the observations do not have infinite spatial resolution and are not free of noise, and therefore, the two intervals with $\beta_0$ and $\beta_8$ are sufficient to explain most of the signal. Moving forward, possible improvements to the modeling approach would be to penalize significant variations of flux between two adjacent bins of $\beta$, assuming that they should be correlated with each other (e.g., \citealp{Jennings2020}), but this out of the scope of the present paper.

\subsubsection{The origin and detectability of the halo}

Regarding the origin of the halo, \citet{Olofsson2022} showed that gas drag can alter the radial distribution of the dust particles (on top of the effect of radiation pressure). Since the disk around HD\,32297 harbors some amount of gas, it begs the question of how this additional drag force impacts our results. But because our modeling approach is focused on the different spatial scales, the exact mechanism responsible for the segregation of the particles does not really matter, whether it is radiation pressure or a combination of gas drag and radiation pressure, as long as the dependence on the size is similar. In both cases, it is the smallest particles that are pushed the farthest away from the star, while the larger grains remain in the birth ring (or even migrate inwards). As a matter of fact, \citet{Olofsson2022} showed that in the presence of gas, the surface brightness of the disk (dominated by the small grains) can only become shallower than in the gas-free case, which can only help creating and maintaining the halo. In that case, the correction factor $\alpha$ would be under-estimated as the small grains can survive longer in the disk compared to the gas-free case. Regardless, even if the segregation of the particles is driven by gas drag and not solely radiation pressure, it supports our conclusion that the halo is caused by small dust particles.

It should be noted though that the detectability of the halo highly depends on the spatial resolution (the beam sizes of the Band\,6 and 8 observations being comparable). While large grains should be confined to the birth ring (a rather compact region), the small grains are set on highly eccentric orbits, and therefore their contribution to the total flux is more spatially diluted. Even if their contribution to the total flux is significantly boosted by the correction factor $[(1-\beta)/(1-2\beta)]^{3/2}$, at higher angular resolution, the contribution of the small particles will nonetheless be spread over several elements of resolution and may not be easily detectable above the noise level (see also, \citealp{Lynch2021}). Finally, it is not surprising that the two disks for which a mm halo has been confirmed (HD\,32297 and HD\,61005, \citealp{MacGregor2018}) are edge-on disks. The edge-on configuration is indeed more favorable to detect such halos as the line of sight goes through more material, even though \citet{Marino2016} also reported the possible detection of a halo (or outer belt) for the face-on disk around HD\,181327.

\subsection{The polarized phase function}\label{sec:pfunc}

\begin{figure}
\centering
\includegraphics[width=\hsize]{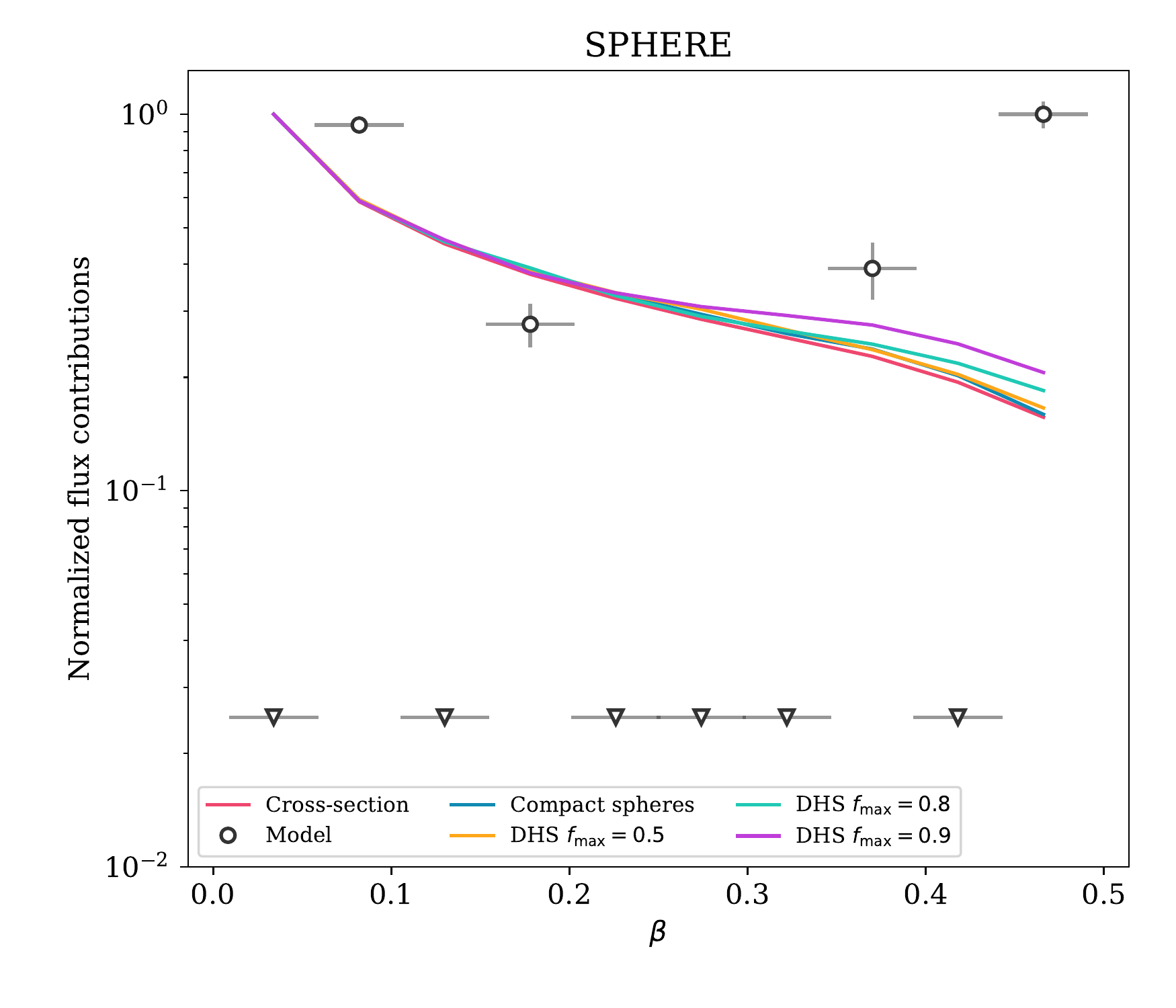}
    \caption{Same as Figure\,\ref{fig:contrib_alma} but for the SPHERE observations}
\label{fig:contrib_sphere}
\end{figure}

\begin{figure}
\centering
\includegraphics[width=\hsize]{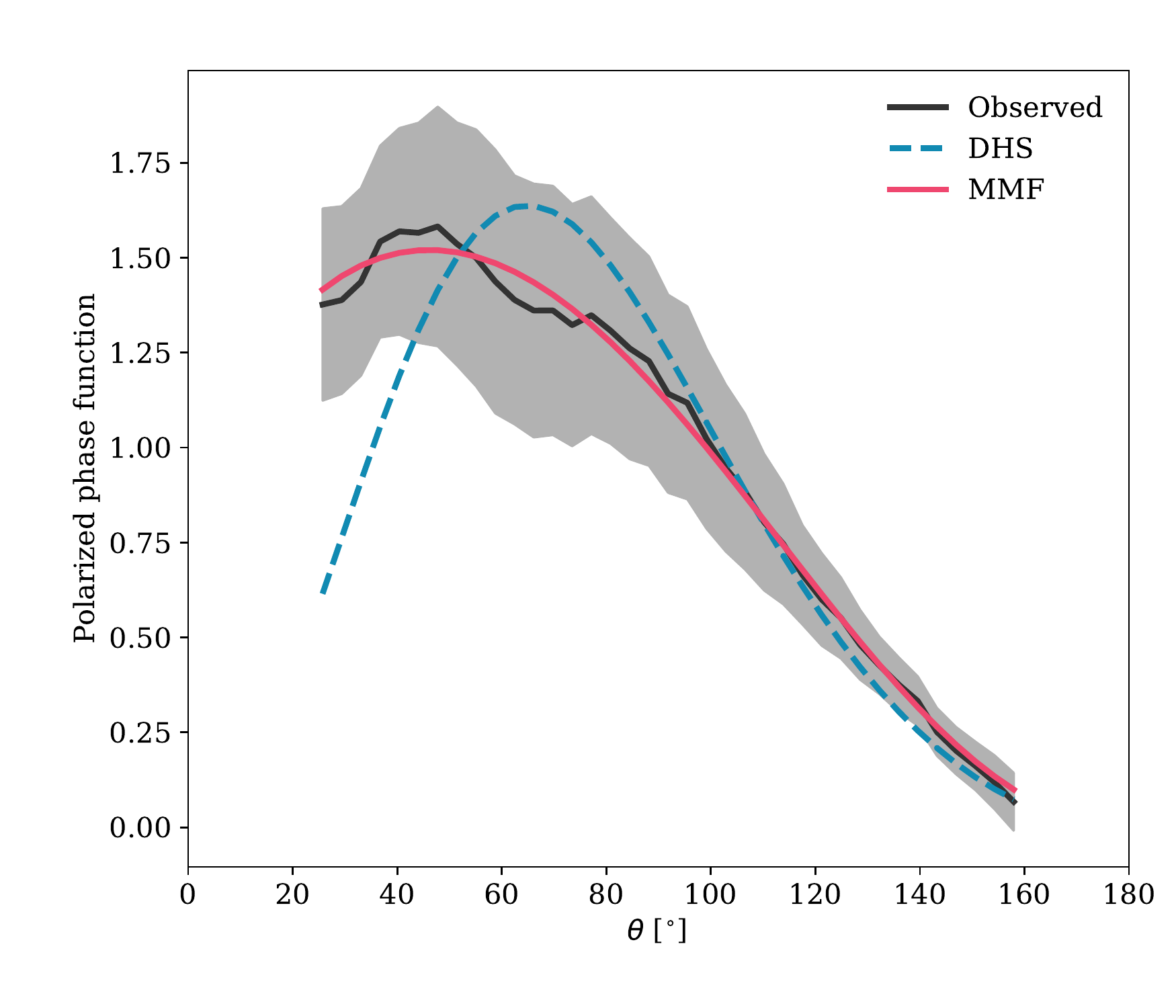}
    \caption{Polarized phase function derived from the modeling of the SPHERE observations (dark line and shaded area), as a function of the scattering angle $\theta$. The best fit solutions using DHS and MMF are shown in dashed blue and solid red, respectively.}
\label{fig:pfunc}
\end{figure}

To model the near-IR SPHERE observations, several $\beta$ intervals are required, ranging between $\sim0.1$ up to $\sim 0.45$, as shown in Figure\,\ref{fig:contrib_sphere}. Similarly to Fig.\,\ref{fig:contrib_alma}, the ``cross-section'' model is shown with a red line, whereas the flux contributions required to fit the SPHERE observations when relaxing the constraint on the size distribution are shown with open black symbols. The other colored lines are the contributions from the ``cross-section'' model, multiplied this time by the absorption efficiencies $Q_\mathrm{sca}$ (also computed using DHS with \texttt{optool}). To successfully reproduce the observations, four intervals of $\beta$ have to contribute above $10$\% of the total flux, $\beta_1 = 0.08$, $\beta_3 = 0.18$, $\beta_7 = 0.37$, and $\beta_9 = 0.47$. In relative terms (since all the profiles are normalized to their maximum), the impact of accounting for $Q_\mathrm{sca}$ is less pronounced than the impact the $Q_\mathrm{abs}$ values had for thermal emission. Nonetheless, departure from spherical compact grains help to increase the contribution of the smallest particles to the total flux in scattered light. That being said, Fig.\,\ref{fig:contrib_sphere} suggests that the models (including scattering efficiencies or not) under-predict the required flux contribution of small dust grains. As mentioned before, this could be related to the use of only a handful of $\beta$ intervals in the modeling, or it could be related to the presence of gas in the system. The ``cross-section'' model does not include the additional effect of outward migration due to the gas drag (which is more efficient on small particles, \citealp{Olofsson2022}), possibly resulting in an under-estimation of the contributions for large $\beta$ values. Lastly, the normalization of the profiles could slightly mislead the interpretation of the Figure. All the profiles for the models are indeed normalized to unity for the interval $\beta_0$ (their maximum), while the contributions for the best fit model are normalized to the last bin $\beta_9$. If instead, all the profiles were normalized to, for instance, the bin $\beta_1$, this will shift down all the open symbols by a factor $\sim 1.8$, bringing the last point closer to the profiles of the models. As mentioned for the modeling of the ALMA observations, introducing a penalty for loosely correlated flux contributions between adjacent bins could be an interesting improvement to the approach.

The polarized phase function derived from the best fit solution is shown as a black solid line in Figure\,\ref{fig:pfunc} and the shaded area correspond to the uncertainties estimated during the fitting process. Even though the disk is almost perfectly edge-on ($i = 86.9^{\circ}$), we cannot probe the full range of scattering angles, since the projected semi-minor axis of the disk lies behind the coronagraph, masking both small and large scattering angles. For this reason, and as noted in \citet{Olofsson2020}, we cannot normalize the phase function over $4\pi$ steradians, and therefore the absolute values of the phase function cannot be properly calibrated. We restrict the range where the shape of the phase function can be trusted to the interval $25^{\circ} \leq \theta \leq 160^{\circ}$. 

We then fitted the polarized phase function using \texttt{optool}, with the same composition as before, using the DHS model with $f_\mathrm{max} = 0.8$. The two free parameters are the size of the grain $s$ (between $0.01$ and $5$\,$\mu$m) as well as the porosity fraction $\mathcal{P}$ (between $0$ and $0.75$, in volume). Since the observed phase function is not normalized over $4\pi$\,steradians, for each model, we find the scaling factor that best minimizes the differences between the model and the observations (see Eq.\,7 of \citealp{Olofsson2016}). Using \texttt{emcee} we found that the best solution is for $s = 0.29\pm0.01$\,$\mu$m and $\mathcal{P} = 0.4\pm0.2$ (mostly unconstrained), and this solution is shown as a blue dashed line in Fig.\,\ref{fig:pfunc}. While the shape at scattering angles larger than $\sim 90^{\circ}$ matches the observed profile relatively well, the best fit solution cannot reproduce the plateau at smaller scattering angles, as the peak of the DHS model is too narrow.

DHS is ideal to compute the optical properties of irregular grains with low porosity. But given that we cannot find a satisfying fit to the observations, we tried another model, the Modified Mean Field Theory (MMF, \citealp{Tazaki2018}), which mimics high porosity grains or aggregates of monomers.  We used the same composition as in Section\,\ref{sec:halo} and for MMF, there are three main parameters; the size of the monomer $s_0$, the size $s$ of the aggregate, and the fractal dimension $D_\mathrm{f}$. For $s_0$ we used the default value of $0.1$\,$\mu$m (e.g., \citealp{Tazaki2016}) and varied $s$ between $0.5$ and $5$\,$\mu$m ($s$ cannot be smaller than $s_0$). The second free parameter is $D_\mathrm{f}$ that we vary between $1.1$ and $2.1$. Unfortunately, there are some limitations when using MMF, especially for small wavelengths and large $D_\mathrm{f}$ values, for which the scattering matrix (as well as the asymmetry parameter) cannot be determined reliably\footnote{We could not use the MMF model for Section\,\ref{sec:halo} and the beginning of this Section for the same reason. We need to compute $g_\mathrm{sca}$ for all the wavelengths of the stellar spectrum to compute $\beta$. Since the stellar model peaks at short wavelengths, we cannot ignore this wavelength range and therefore cannot properly estimate $\beta$.} as multiple scattering events cannot be neglected. We therefore could not explore the range $2.1 < D_\mathrm{f} \leq 3$. As noted in \citet{Tazaki2021}, the range of $D_\mathrm{f}$ between $1.7-2.1$ would correspond to fluffy aggregates formed by ballistic cluster cluster aggregation (BCCA), and $D_\mathrm{f}$ can be in the range $1.1-1.4$ for non-ballistic CCA clusters (analogous to a linear chain of monomers). Even though we could not compute the phase function for those cases, larger values of $D_\mathrm{f}$ ($\sim 3$) would correspond to a ballistic particle cluster aggregation process (BPCA), leading to nearly homogeneous aggregates with high porosity (\citealp{Tazaki2021} and references therein). Using \texttt{emcee}, we found that the best solution is obtained for $D_\mathrm{f} = 1.73 \pm 0.01$ and $s \geq 2.5$\,$\mu$m, and the corresponding polarized phase function is shown as a solid red line in Fig.\,\ref{fig:pfunc}. The range of possible values for $D_\mathrm{f}$ is narrow, but our best fit value for $s$ is a lower limit, as grains larger than $2.5$\,$\mu$m also provide an equally good fit to the derived phase function. The value $D_\mathrm{f} \sim 1.7$ suggests a ``fluffy'' structure for the aggregates, which should have been formed by collisions between clusters of comparable sizes (\citealp{Tazaki2016}). 

To better understand the origin of those fluffly aggregates we can turn to studies of comets in the solar system. As outlined in the review by \citet{Levasseur-Regourd2018}, the current paradigm is that cometary dust is mostly composed of aggregates of various ``compactness''. Most recently, the Rosetta mission provided valuable insights on the constituents of the 67P/Churyumov-Gerasimenko (67P hereafter) comet. Analyzing data from the Grain Impact Analyzer and Dust Accumulator (\citealp{Colangeli2007}) instrument, \citet{Fulle2015} identified two different populations of particles; compact particles and fluffy aggregates, the latter having filling factors (the fraction of the particle volume occupied by monomers) as low as $10^{-3}$. A further analysis of the data by \citet{Fulle2016} predicted that the fractal dimension of the aggregates should be close to or smaller than $\sim 1.87$, and that those particles should amount to $\sim 15$\% of the non-volatile volume. In parallel, \citet{Mannel2016} analyzed observations taken by the Micro-Imaging Dust Analysis System (\citealp{Riedler2007}) on board the Rosetta spacecraft, and even though most of the particles analyzed are quite compact, they detected one fluffy aggregate with an estimated fractal dimension of $D_\mathrm{f} = 1.7$.

Our current understanding of grain growth is that fluffy aggregates must have formed during the proto-planetary phase. The small relative velocities of the sub-$\mu$m-sized monomers leads to the formation of aggregates, whose fractal dimension is smaller than $2$ (see \citealp{Blum2008} for a review). Grain growth continues in a ``hit-and-stick'' regime until we reach the bouncing barrier, which should happen for sizes between $1$\,cm to $1$\,mm depending on the location in the disk (\citealp{Zsom2010}, \citealp{Lorek2018}), leading to the compaction of the particles. Consequently, fractal particles must be pristine, and must have survived the compaction phase during the formation of planetesimals. Such particles have most likely been incorporated within the voids between larger pebbles when forming larger bodies. Based on the Rosetta mission observations, \citet{Fulle2017} suggested that 67P cannot have experienced any catastrophic collision in its lifetime, which would have otherwise resulted in compaction and destruction of the fluffy aggregates. The age of HD\,32297 is not well determined, but the star should be older than $15$\,Myr (\citealp{Rodigas2014}) but younger than $\sim 30$\,Myr (\citealp{Kalas2005}), a possible analog to the young solar system. It is therefore \textit{plausible} that the second dust generation we observe around HD\,32297 is the result of collisions of the progenitors of cometary bodies similar to 67P. They would contain a fraction of pristine fractal aggregates within their nuclei, and may have evolved over several Myr without suffering any major collisions (thus avoiding compaction of the aggregates), until they are eventually destroyed to release the particles that we observe. The smallest of those particles are then set on high eccentricity orbit where they can survive for long periods of time. Nonetheless, this explanation is not free of hurdles; \citet{Thebault2007} found that for a disk as luminous as the one around HD\,32297, the collisional timescale of the particles should be much shorter than the age of the star. It is therefore a timescale that is hard to reconcile with the possible pristine nature of the particles. Measuring the total intensity phase function and deriving the degree of polarization (the ratio between polarized and total intensity) would greatly help confirming that the particles are indeed akin to fluffy aggregates. Unfortunately, we cannot readily use the angular differential imaging observations presented in \citet{Bhowmik2019} as the post-processing of the observations strongly biases the determination of the total intensity phase function, especially near the projected minor axis of the disk. Alternative post-processing techniques, such as reference star differential imaging, making use of a calibration star, might alleviate some of those issues.

It should also be noted though that the typical sizes of the particles detected in the vicinity of 67P are smaller than $\sim 1$\,mm, larger than the size of the aggregates required to reproduce the polarized phase function of HD\,32297 (even though our best fit value is a lower limit). Measuring the degree of polarization of the debris disk would provide additional constraints on the properties of the dust particles. As a final remark, as mentioned previously, the modeling approach assumes that the phase function is the same for all the particles, regardless of their sizes, a strong but necessary assumption. Additional observations, both in total intensity and polarization, at optical and near-IR wavelengths (other than $J$ band), might help in that regard.

\section{Summary}

In this paper we presented a new approach to model multi-wavelength spatially resolved observations of debris disks. The emphasis is put on the spatial extent of particles that experience different strength of the radiation pressure force. By parametrizing the model using different intervals of $\beta$ values, we are able to alleviate the modeling of the relationship between the grain size and $\beta$ as much as possible (but not entirely). The characterization of the optical properties can then be done a posteriori.

By modeling simultaneously near-IR SPHERE and mm ALMA observations of the disk around HD\,32297, we take advantage of the synergy between the different wavelengths, better constraining the location of the parent planetesimals to model both datasets. We find that the disk is best described by a reference radius of $1\arcsec$ and a standard deviation for the width of $0.22\arcsec$ and the disk is not resolved vertically in either dataset. The ALMA observations are best reproduced by a combination of two intervals of $\beta$ values (small and large $\beta$), while the SPHERE observations require several intervals of $\beta$ values to be fitted adequately.

We confirm the presence of extended emission in the ALMA Band\,8 observations, corresponding to the halo reported in \citet{MacGregor2018}. We show that this halo does not necessarily originate from the thermal emission of large mm-sized grains (which would require a mechanism to excite their eccentricities above $0.7$), but rather that small grains just above the blow-out size can explain this extended emission (that amounts to about $\sim 10$\% of the total flux). This is due to the larger number of such particles, in part because of the top-heavy size distribution, but most importantly, because of their eccentric orbits significantly increasing their lifetime. Both effects contribute to compensating for their poorer emissivity at long wavelengths. Because of the large eccentricity involved, the detectability of the halo should also depend on the spatial resolution of the observations, making it more difficult to detect at larger resolution. We also discuss the fact that irregularly shaped grains can increase the total flux by an order of magnitude compared to compact spherical grains, for the same total dust mass. However, even for irregular grain shapes, models that include the dust opacity values $Q_\mathrm{abs}$ are still a factor $2-3$ below the measured $\sim 10$\% contribution of the halo to the total flux in the mm, which thus remains an open question to be further investigated. 

To model the polarized phase function determined from the SPHERE observations, the Modified Mean Field Theory model provides the most satisfactory fit to the observations. We find that it is best modeled by aggregates with a size larger than $2.5$\,$\mu$m, with a fractal dimension of $1.7$, suggesting that the particles we detect at near-IR wavelengths are comparable to fluffy aggregates. Such particles have been detected in comets of the solar system (e.g., 67P) and are considered to be pristine remnants of its infancy. We hypothesize that the particles in the disk around HD\,32297 were released from the collisions of similar cometary bodies, that have not suffered many collisions before being eventually destroyed.

\begin{acknowledgements}
We thank the referee for their helpful comments, that helped clarify several parts of the paper, especially regarding the assumptions that are implicitly made in the modeling approach.
J.\,O. and A.\,B. acknowledge support by ANID, -- Millennium Science Initiative Program -- NCN19\_171. J.\,O. acknowledges support from the Universidad de Valpara\'iso, and from Fondecyt (grant 1180395). 
G.\,M.\,K. is supported by the Royal Society as a Royal Society University Research Fellow.
This paper makes use of the following ALMA data: ADS/JAO.ALMA\#2017.1.00201.S. ALMA is a partnership of ESO (representing its member states), NSF (USA) and NINS (Japan), together with NRC (Canada), MOST and ASIAA (Taiwan), and KASI (Republic of Korea), in cooperation with the Republic of Chile. The Joint ALMA Observatory is operated by ESO, AUI/NRAO and NAOJ.
SPHERE is an instrument designed and built by a consortium consisting of IPAG (Grenoble, France), MPIA (Heidelberg, Germany), LAM (Marseille, France), LESIA (Paris, France), Laboratoire Lagrange (Nice, France), INAF–Osservatorio di Padova (Italy), Observatoire de Gen\`eve (Switzerland), ETH Zurich (Switzerland), NOVA (Netherlands), ONERA (France) and ASTRON (Netherlands) in collaboration with ESO. SPHERE was funded by ESO, with additional contributions from CNRS (France), MPIA (Germany), INAF (Italy), FINES (Switzerland) and NOVA (Netherlands).  SPHERE also received funding from the European Commission Sixth and Seventh Framework Programmes as part of the Optical Infrared Coordination Network for Astronomy (OPTICON) under grant number RII3-Ct-2004-001566 for FP6 (2004–2008), grant number 226604 for FP7 (2009–2012) and grant number 312430 for FP7 (2013–2016). We also acknowledge financial support from the Programme National de Plan\'etologie (PNP) and the Programme National de Physique Stellaire (PNPS) of CNRS-INSU in France. This work has also been supported by a grant from the French Labex OSUG@2020 (Investissements d'avenir – ANR10 LABX56). The project is supported by CNRS, by the Agence Nationale de la Recherche (ANR-14-CE33-0018). It has also been carried out within the frame of the National Centre for Competence in Research PlanetS supported by the Swiss National Science Foundation (SNSF). MRM, HMS, and SD are pleased to acknowledge this financial support of the SNSF. 
This work has made use of data from the European Space Agency (ESA) mission {\it Gaia} (\url{https://www.cosmos.esa.int/gaia}), processed by the {\it Gaia} Data Processing and Analysis Consortium (DPAC, \url{https://www.cosmos.esa.int/web/gaia/dpac/consortium}). Funding for the DPAC has been provided by national institutions, in particular the institutions participating in the {\it Gaia} Multilateral Agreement.
This research made use of Astropy,\footnote{\url{http://www.astropy.org}} a community-developed core Python package for Astronomy \citep{astropy:2013, astropy:2018}, Numpy (\citealp{numpy}), Matplotlib (\citealp{matplotlib}), Scipy (\citealp{scipy}), and Numba (\citealp{numba}).
\end{acknowledgements}

\bibliographystyle{aa}

\appendix

\section{Corner plots}

\begin{figure*}
\centering
\includegraphics[width=0.9\hsize]{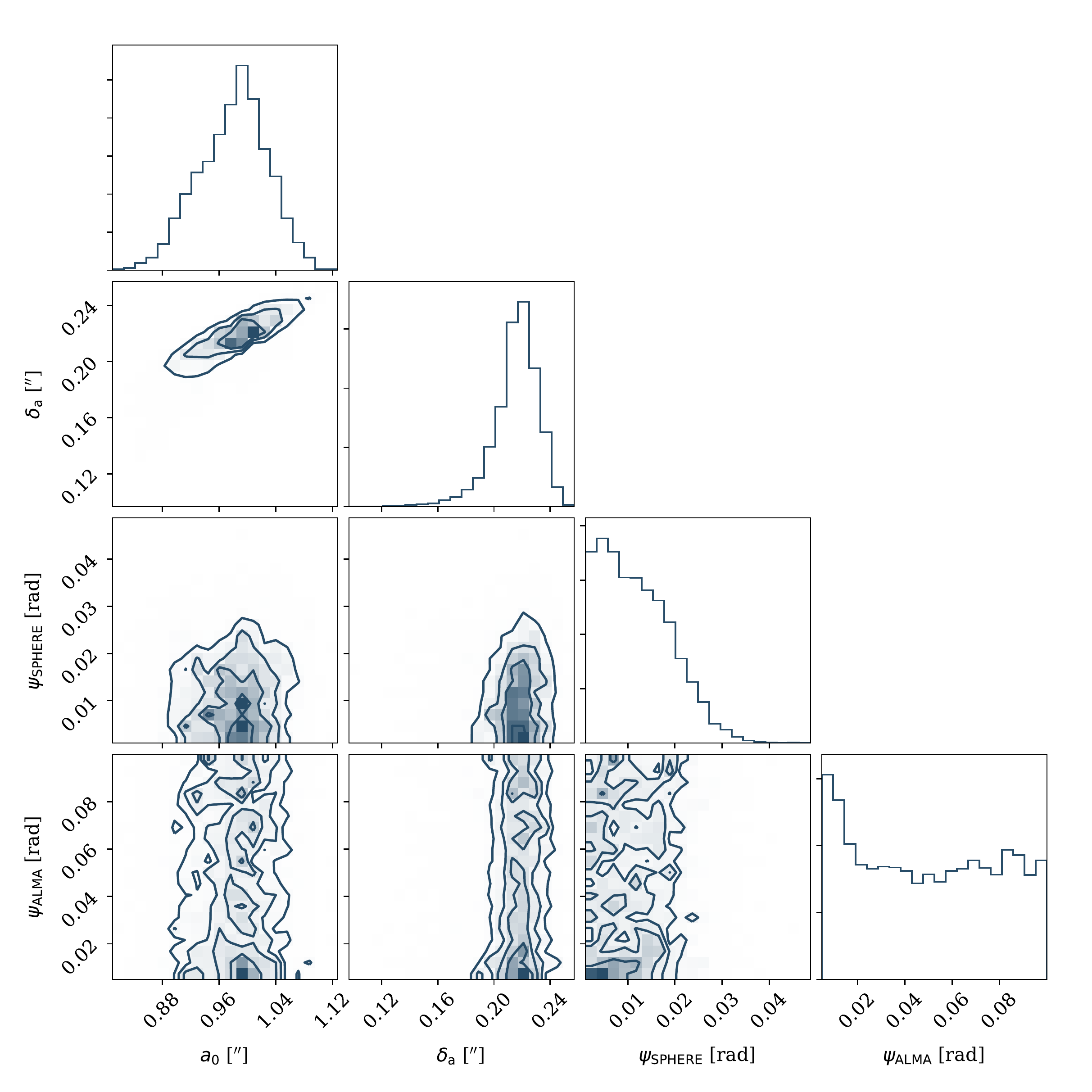}
    \caption{Corner plot for the modeling results of the SPHERE and ALMA observations}
\label{fig:corner_obs}
\end{figure*}

Figure\,\ref{fig:corner_obs} shows the 2D density distributions as well as the projected probability density distributions, for the free parameters when modeling the SPHERE and ALMA observations.

\end{document}